\documentclass[conference]{IEEEtran}
\IEEEoverridecommandlockouts
\usepackage{cite}
\usepackage{amsmath,amssymb,amsfonts}
\usepackage{algorithmic}
\usepackage{graphicx}
\usepackage{textcomp}
\usepackage[acronym]{glossaries}

\newacronym{WTA}{WTA}{Winner-Take-All}
\newacronym{PSP}{PSP}{Post Synaptic Potential}
\newacronym{XOR}{XOR}{Exclusive Or}
\newacronym{QUBO}{QUBO}{Quadratic Unconstrained Binary Optimization}
\newacronym{STDP}{STDP}{Spike Time Dependent Plasticity}
\newacronym{ANN}{ANN}{Artificial Neural Network}
\newacronym{SNN}{SNN}{Spiking Neural Network}
\newacronym{SSSP}{SSSP}{Singe Source Shortest Path}
\newacronym{LCA}{LCA}{Local Competitive Algorithm}
\newacronym{LIF}{LIF}{Leaky Integrate and Fire}
\newacronym{CMOS}{CMOS}{Complementary Metal-Oxide-Semiconductor}
\newacronym{VLSI}{VLSI}{Very Large Scale Integration}
\newacronym{NoC}{NoC}{Network on Chip}
\newacronym{SRM}{SRM}{Spike Response Model}
\newacronym{QA}{QA}{Quantum Annealing}
\newacronym{AER}{AER}{Addressed Events Representation}
\newacronym{FPGA}{FPGA}{Field Programmable Gate Array}
\newacronym{SIMD}{SIMD}{Single Instruction Multiple Data}
\newacronym{ASIC}{ASIC}{Application-Specific Integrated Circuit}
\newacronym{ISA}{ISA}{Instruction Set Architecture}
\newacronym{MSE}{MSE}{Mean Squared Error}
\newacronym{MCMC}{MCMC}{Markov Chain Monte Carlo}
\newacronym{IO}{IO}{Input/Output}
\newacronym{AdEx}{AdEx}{Adaptive Exponential}
\newacronym{NLIF}{NLIF}{Non-linear Integrate-and-Fire}
\newacronym{IF-SFA}{IF-SFA}{Integrate-and-Fire with Spike
Frequency Adaptation}
\newacronym{SpMV}{SpMV}{Sparse Matrix Vector Product}
\newacronym{SA}{SA}{Simulated Annealing}
\newacronym{TSP}{TSP}{Traveling Salesman Problem}
\newacronym{SNP}{SNP}{Single Nucleotide Polymorphisms}
\newacronym{GWAS}{GWAS}{Genomic-Wide Association Studies}
\newacronym{DNA}{DNA}{Deoxyribonucleic Acid}
\newacronym{BIC}{BIC}{Bayesian Information Criterion}
\newacronym{POPC}{POPC}{Population Count}
\newacronym{GPU}{GPU}{Graphical Processing Unit}
\usepackage{xcolor}

\def\BibTeX{{\rm B\kern-.05em{\sc i\kern-.025em b}\kern-.08em
    T\kern-.1667em\lower.7ex\hbox{E}\kern-.125emX}}
\begin{document}

\title{Constant Depth Threshold Circuits For Exhaustive Epistasis Detection\\

}

\author{\IEEEauthorblockN{
André Ribeiro}
\IEEEauthorblockA{
\textit{INESC-ID, Instituto Superior Técnico} \\ Universidade de Lisboa, Portugal \\
andre.a.ribeiro@tecnico.ulisboa.pt
}
\and
\IEEEauthorblockN{
Aleksandar Ilic}
\IEEEauthorblockA{
\textit{INESC-ID, Instituto Superior Técnico} \\ Universidade de Lisboa, Portugal \\
aleksandar.ilic@inesc-id.pt
}
\and
\IEEEauthorblockN{
Leonel Sousa}
\IEEEauthorblockA{
\textit{INESC-ID, Instituto Superior Técnico} \\  Universidade de Lisboa, Portugal \\
leonel.sousa@inesc-id.pt
}}

\maketitle

\begin{abstract}

The development of large-scale neuromorphic hardware has made practical implementations of threshold gate-based circuits a near-term possibility. The complexity advantages regarding traditional computing classes, as evidenced in the literature, have prompted us to tackle Epistasis Detection, one of the most computationally complex combinatorial problems in bio-informatics. We propose specially designed circuits that calculate the relative frequencies of all dataset combinations in an efficient pipelined fashion, taking advantage of co-located memory and configurable parallelism, obtaining complexity gains. Overall, we obtain the runtime to be bounded by the number of combinations to calculate, without any additional complexity overhead, contrary to classical approaches, using log-linear space. To accomplish this, we propose a data encoding and combination generation strategy using \gls{LIF} neurons, that feeds a constant depth threshold gate population count circuit. Accounting for typical hardware characteristics, such as limited fan-in and variable precisions, we obtain logarithmic depth and log-cubic linear connections, for the population count circuit by composing developed unbounded fan-in constant depth threshold gate circuits to perform population count and binary array sum.

\end{abstract}

\begin{IEEEkeywords}
Epistasis Detection, Threshold Gates, Complexity Theory, Population Count, Binary Sum, Synaptic Memory.
\end{IEEEkeywords}

\section{Introduction}

Neuromorphic computing, inspired by the structure of the brain, has emerged as a promising alternative to conventional computing. It offers advantages such as inherent parallelism, co-located processing and memory, mitigating the von Neumann bottleneck, and energy efficiency, as computation occurs only when neurons are active \cite{aimone_provable_2021}. Recent advancements in large-scale neuromorphic hardware have made the practical implementation of threshold gate circuits a near-term possibility, which are more powerful than classical boolean circuits  \cite{parekh2018constant, aimone_provable_2021}. Prior work has demonstrated complexity advantages for specific problems such as polynomial improvements in graph algorithms \cite{aimone_provable_2021}, sub-cubic space constant-depth matrix multiplication \cite{parekh2018constant}, and minimum spanning tree detection \cite{janssen2024solvingMST}.

While neuromorphic systems gain track in edge applications, their potential for processing large datasets remains a challenge \cite{schuman_opportunities_2022}. For this reason, theoretical approaches must consider hardware constraints, such as limited fan-in and variable precision, as ignoring them may lead to unrealistic conclusions. For instance, assuming rational-valued currents allows to encode infinite information in a single neuron using Cantor-like stack encoding \cite{siegelmann1992computational}, while unbounded fan-in enables theoretically possible, though practically unachievable, constant-depth circuits \cite{TC_bounded_fan}.

In this work, we step toward enabling neuromorphic computing for large dataset processing. We develop a full neuromorphic circuit that leverages co-located memory and threshold gate circuits computational advantages to address epistasis detection; one of the most computationally demanding combinatorial problems in bio-informatics today \cite{gracca2024distributedILLIC}. Epistasis detection involves correlating phenotypes, such as disease presence, with the frequency of specific \glspl{SNP} at different DNA positions across samples. Accurate capture requires calculating frequencies for all possible combinations of positions and polymorphisms, which becomes computationally prohibitive as dataset size and interactions grow \cite{nobre2021fourthILLIC}. 

Though it is possible to lower the computational burden, with strategies such as pre-filtering \cite{NielEPISTASIS} and state-of-the-art transformer-based approaches \cite{gracca2024distributedILLIC}, exhaustive epistasis search remains indispensable for uncovering all relevant genetic interactions \cite{nobre2021fourthILLIC}. For all possible combinations of DNA positions, and \glspl{SNP} values in the selected positions, a mask of the number of samples is computed by a vectorized AND, and frequencies are counted using \gls{POPC}.  The most performant approaches rely on \gls{GPU}, incurring high power consumption, requiring many memory transfers, and synchronization \cite{nobre2021fourthILLIC}.

We propose a neuromorphic approach to exhaustive epistasis detection by designing circuits that construct a contingency table in a fully pipelined manner, without the need for off-chip memory transfers, and developing efficient depth \gls{POPC} circuits to build the required contingency tables. Our method accounts for typical neuromorphic hardware constraints while maintaining low time complexity. To our knowledge, this is the first effort to provide a neuromorphic computing algorithm for Epistasis detection. Specifically, this paper proposes:
\begin{itemize} 
    \item a \gls{POPC} and binary sum circuit with constant depth; 
    \item adaptations of these circuits to logarithmic depth under fan-in and variable precision constraints; 
    \item an addressable, non-destructive encoding of firing patterns using stack encoding; 
    \item a neuromorphic circuit to compute \gls{SNP} combination frequencies for exhaustive epistasis detection. 
\end{itemize}

This paper is organized as follows. In Section \ref{sec:background} we present the relevant background network model. In Section \ref{sec: Epistasis} we provide a high-level description of the proposed circuit and introduce hardware restrictions. In section \ref{sec:Seq encoding and Gen} we propose a dataset encoding scheme. In Section \ref{sec:POPC_general} we propose the \gls{POPC} circuit. Finally, in Section \ref{sec:discussion} we present the final computational complexity results and conclude in \ref{sec:conclusions} along with future directions of investigation.

\section{Background}
\label{sec:background}

We provide an Epistasis formulation and introduce our neuromorphic circuit model based on \cite{cp2020power}. We also present a classical threshold gate circuit to compute the parity of a string of bits in constant depth, linear space, and quadratic connections, that we will resort to for the \gls{POPC} circuit.

\subsection{Exhaustive Epistasis Detection}
\label{sec:enunciating Epistsais}

Epistasis detection deals with the identification of interactions of \glspl{SNP} that are responsible for a phenotypic trait, e.g. a particular characteristic or disease. In computational terms, epistasis algorithms tackle the detection problem by processing genotype information from a dataset $D$ of size $M \times (N + 1)$; $M$ refers to the number of patient records or samples (in the case of disease traits) and $N$ is the number of \glspl{SNP} to be studied. Each entry $D[i, j]$, $i \in \{1, ..., M \}$, $j \in \{1, ..., N \}$, represents the genotypic value at the $j$-th \gls{SNP} from the $i$-th sample. The value stored in $D[i, j]$ depends on the allele configuration and can take the values 0 (homozygous major allele), 1 (heterozygous allele), or 2 (homozygous minor allele). The trait observed at the $i$-th record, which is stored in $D[i, N + 1]$, is set to 0 if $i$ is a control or 1 if $i$ is a case. Assuming an interaction order $k$, a candidate interaction is encoded by a combination of $k$ \glspl{SNP} $X = [X_1, X_2, ..., X_k]$, where the alleleic information $X_i \in \{1, ..., N \}$. High-order epistasis detection is a very challenging task, due to the large number of sets of \glspl{SNP} to evaluate, $\binom{M}{k} = \frac{M!}{k!(M - k)!}$, which grows exponentially with $k$. In practice, $k$-order Epistasis detection is performed for values of $k$  between $2$ and $4$ \cite{nobre2021fourthILLIC}.

\begin{figure}[t]
    \centering
    \includegraphics[width=1\linewidth]{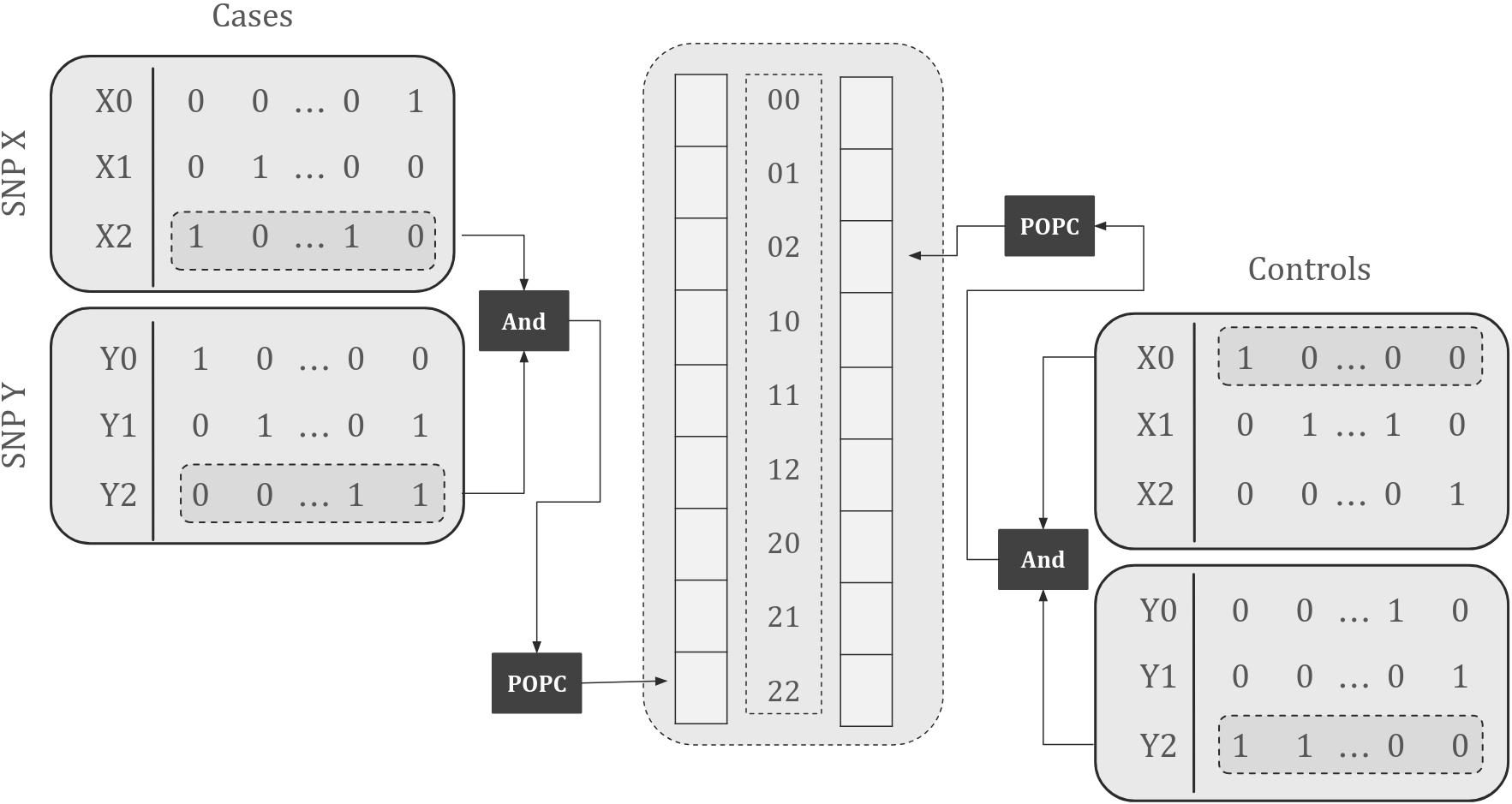}
    \caption{Contingency table construction example for
a pairwise interaction using a binarized data representation (adapted from \cite{nobre2021fourthILLIC}).}
    \label{fig:nobreTable}
\end{figure}

To enable an exhaustive search, the first step is to construct a contingency table that separately records the frequency of each \gls{SNP} combination among control and case samples. Since each \gls{SNP} can take one of three values, a $k$-SNP interaction results in $3^k$ possible combinations, requiring the construction of a table with $3^k$ entries for each of the $\binom{M}{k}$ combinations. To reduce computational overhead, the approach from \cite{nobre2021fourthILLIC} encodes each \gls{SNP} using a 3-bit masked one-hot format: ($0 \rightarrow 100$, $1 \rightarrow 010$, $2 \rightarrow 001$). This encoding allows the efficient mapping of the contingency table construction to binary AND and \gls{POPC} operations, being amenable to a spike representation. The procedure is illustrated in Figure \ref{fig:nobreTable} for $k=2$, using \glspl{SNP} $X$ and $Y$. Each table entry is computed using a \gls{POPC} operation over the samples matching the corresponding \gls{SNP} combination. Once built, statistical tests such as the $\chi^2$ test or mutual information are applied to evaluate the significance of detected interactions \cite{NielEPISTASIS}. In this work, we focus on building the contingency table, since it represents the main computational challenge \cite{nobre2021fourthILLIC}.

\subsection{Spiking Neural Network Model}
\label{sec:model}
For our spiking neural network, we use a neuron model similar to \cite{cp2020power}. It defines a discrete-timed spiking neural network as a labeled finite digraph comprised of a set of neurons $N$ as vertices and a set of synapses $S$ as directed edges. Every neuron $j$ is a triple $(T_j, m_j, b_j)$ representing the threshold, leakage constant, and bias,  while a synapse $s \in S$ is a 4-tuple $(j, l, d, w )$ for the pre-synaptic neuron, post-synaptic neuron, synaptic delay and weight, respectively.  

The basic principle is thus that any spike of a neuron $j$ at timestep $t$, $x_j(t)$ is carried along outgoing synapses with weight $w_{jl}$ to serve as inputs to the receiving neuron's $l$ current, $u_l$, at timestep $t+1$. The behavior of a neuron at time $t+1$ is defined using its membrane potential 
\begin{equation}
    u_j(t+1) = m_j u_j(t) + b_j + \sum_{k} w_{kj} x_k(t-d_{kj})
\end{equation} 
which is the integrated weighted sum of the neuron's inputs (taking into account synaptic delay) plus an additional bias term. Whether a neuron spikes or not at any given time is dependent on its membrane potential being \textit{strictly greater} than the threshold. If there is a spike, the current is reset to $0$. A spike $x_k(t)$ is abstracted here to be a singular discrete event, i.e., $x_k(t) = 1$ if a spike is released by neuron $x_k$ at time $t$ and $x_k(t) = 0$ otherwise. The neuron's spikes can also be defined programmatically rather than through its membrane potential, with predetermined spiking schedules \cite{cp2020power}.  We also use the in-degree (resp. out-degree) to refer to the number of a neuron's incoming (resp. outgoing) synapses.

\begin{figure}[t]
    \centering
    \includegraphics[width=1\linewidth]{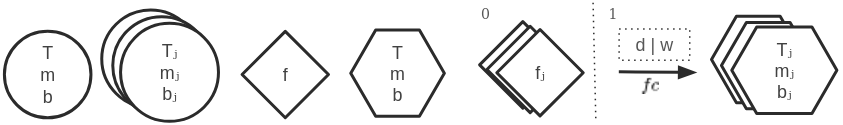}
    \caption{Notational convention for (from left to right): a regular neuron with threshold $T$ leakage $m$ and bias $b$, a layer of regular neurons indexed by $j \geq 1$, a programmed neuron according to a function $f$, a read-out neuron, and fully connected synapses between a programmed neurons layer and a readout layer (adapted from \cite{janssen2024solvingMST}).}
    \label{fig:neuronchematic}
\end{figure}

To represent the neuromorphic circuit we resort to the scheme as presented in \cite{janssen2024solvingMST}. In Figure \ref{fig:neuronchematic}, we introduce the notational convention for graphically depicting spiking neural networks. Note that we make a distinction between regular neurons used for internal computation and readout neurons, for which we can observe the spiking events. When a layer is denoted, the parameters are indexed by $j \geq 1$. Differently from \cite{janssen2024solvingMST}, we denote that programmed neurons fire according to a function $f(t)$ that associates each time step with the neuron's firing behavior. We also denote the threshold $T$, leakage $m$, and bias $b$. Synapses are represented as arrows between neurons. Arrows between layers do not have predefined connection scheme, unless the symbol $fc$ is placed, representing a fully connected synaptic scheme. Dotted lines split depth levels, which are numbered at the top left of each split. The composition of circuits can by performed by switching the outputs with the input neurons of the other.

To compare the efficiency of our proposed spiking implementation, we perform a complexity analysis using measures tailored for neuromorphic algorithms, as proposed in \cite{cp2020power} and followed in \cite{janssen2024solvingMST}, evaluating SNN performance based on time complexity (number of timesteps), space complexity (number of neurons and synapses), and energy complexity (number of spikes). A comparison between neuromorphic and classical hardware, can be enhanced by accounting for data-movement costs, related to spike delivery, that are dependent on the circuit hardware mapping \cite{aimone_provable_2021}, which we do not yet consider.

\subsection{Threshold Gates Class}

The model presented in Section \ref{sec:model} is an adaptation from widely used \gls{LIF} neurons \cite{janssen2024solvingMST}. These networks are more general than feed-forward threshold gate circuits, as the neuron behavior has state variables (e.g. current) that evolve, and allow for recurrent connections. Threshold gate circuits can be implemented as \gls{LIF} neurons with leakage $m$ set to zero. By restricting connections to be feedforward, any neuromorphic computer that allows \gls{LIF} models such as Loihi \cite{Loihi1}, can also implement threshold gate circuits. Moreover, the schematics in Figure \ref{fig:neuronchematic} allow to specify threshold gate circuits in a ready-to-implement fashion. Furthermore, in feedforward circuits, neurons at depth $i$ depend solely on neurons at depth $i-1$ which enables optimally pipelined behavior. The complexity class of constant depth threshold gate circuits is named $TC^0$, which holds more circuits than traditional computation classes, such has boolean circuit ($AC^0$) \cite{aimone_provable_2021}.

\subsection{Constant Depth Parity Circuit}
\label{sec:Constant Depth XOR}

\begin{figure}[t]
    \centering
    \includegraphics[width=0.85\linewidth]{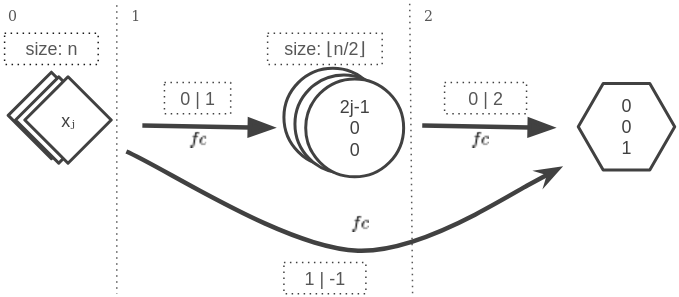}
    \caption{Threshold gate circuit to compute parity of a bit-string.}
    \label{fig:TC_PARITY}
\end{figure}

Our proposed \gls{POPC} circuit relies on the classical \textit{TC} parity (or XOR) circuit. The general idea to compute the parity of a bit-string is to check if the sum of the bits is divisible by two, i.e. if the sum $y$ (irrespective of the base) is even if and only if $\lfloor y/2 \rfloor \times 2 = y$, following the approach presented in \cite{paturi1990thresholdPARITY}. The circuit checks this condition using the two levels of depth represented in Figure \ref{fig:TC_PARITY}. Let $(y_1, y_2, ..., y_n)$ be a bit string such that $\sum_{i=1}^n y_i= y$ encoded in the spikes of the input layer by assigning a neuron to each bit and representing the its value, 1 or 0, by a spike or absence of a spike, respectively.

The first level computes a representation of 
\begin{equation}
\lfloor (\sum_{i=1}^n y_i/2) \rfloor
\end{equation}
It is composed of $\left \lfloor \frac{n}{2} \right \rfloor $
neurons with odd thresholds $t \in \{ 1, 3, 5 , ..., 2\lfloor n/2 \rfloor-1$ \}, such that neuron $j$'s threshold is $2j-1$. Every depth 1 neuron is connected to each input by synapses with weight one. When the input layer spikes, the current induced in every neuron will be equally set to $\sum_{i=1}^n y_i$. Thus, the number of neurons to spike will exactly be $\lfloor (\sum_{i=1}^n y_i/2 \rfloor$, because how the thresholds were set.

The second level consists of a single neuron that checks if 
\begin{equation}
\label{eq:mathcXOR}
    \left\lfloor \frac{\sum_{i=1}^n y_i}{2} \right\rfloor \times 2 = \sum_{i=1}^n y_i,
\end{equation}
by having its threshold set to $0$ and bias $1$. It will receive the sum of the number of spikes of the depth 1 layer scaled by $2$. It will also receive the sum of the input bits scaled by $-1$, as the synapses connecting the input will only deliver their spikes after holding for one timestep. The induced current can only be $0$ or $-1$. If it is 0, then both sides of \eqref{eq:mathcXOR} match and the bit-string has an even number of $1$'s. The bias will take care of pushing the current over the threshold to produce a spike. If the current is $-1$, the bit-string has an odd number of $1$'s,  the bias is not sufficient to produce a spike.   

\subsubsection{Complexity Analysis}
As can be inferred by Figure \ref{fig:TC_PARITY}, the circuit depth does not depend on input size and thus it has $O(1)$ time complexity. In terms of space, it has 
$O(n)$
neurons and, as all the layers are fully connected, it has
$O(n^2)$
synapses. The maximum in-degree is in the output neuron, having $ n +\lfloor \frac{n}{2} \rfloor$ incoming synapses. The maximum out-degree is $n+1$ in the input layer. The maximum synaptic delay is $1$ and is not dependent on the input size. Finally, the energy complexity depends on the actual value of the input and grows with the number of input spikes it receives. It is bounded by all the inputs spiking, leading to $O(n)$ spikes. 

\section{Proposed Epistasis Detection Circuit}
\label{sec: Epistasis}

We focus on building a contingency table for $k$-order epistasis detection, by using the exhaustive search approach presented in Section \ref{sec:enunciating Epistsais}. The objective is to calculate the frequency of all \gls{SNP} combinations across samples. The dataset structure places binarized \gls{SNP} representations as rows and samples as columns, with $3^k$ possible genotypes to compute for each \gls{SNP} subset. To compute a table entry, we count the number of columns where selected \glspl{SNP} both have their value set $1$ as in Figure \ref{fig:nobreTable}. We compute the table for cases and controls in entirely separate runs, by splitting the dataset by the corresponding case and control columns.

\begin{figure}[t]
    \centering
    \includegraphics[width=0.9\linewidth]{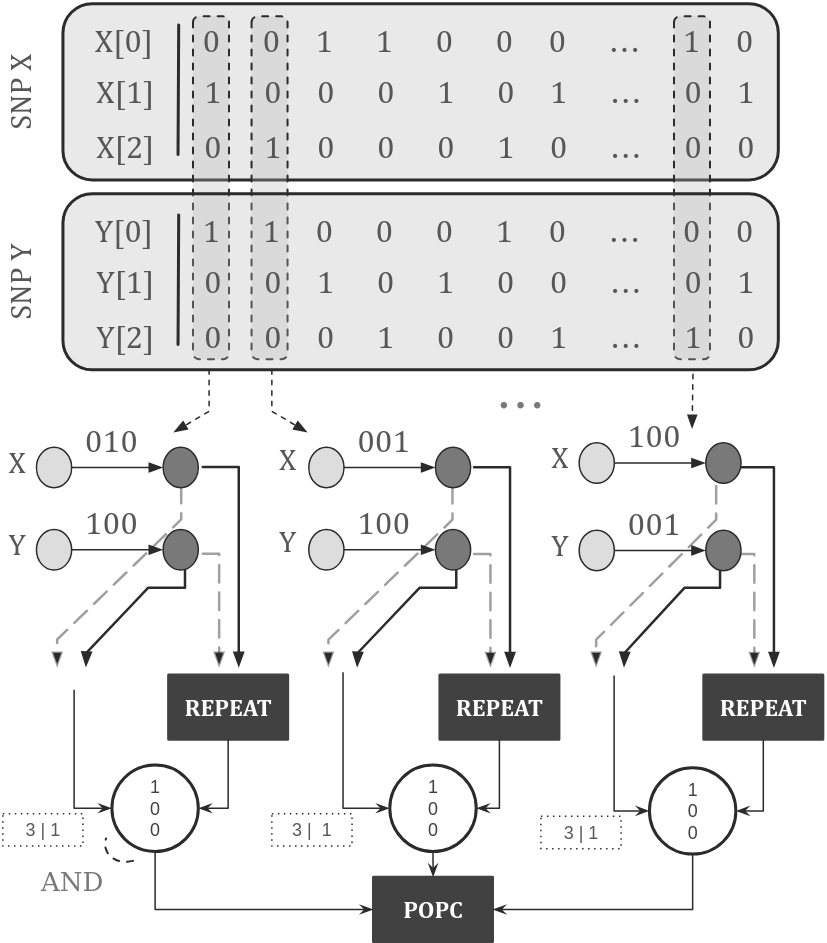}
    \caption{Circuit schematization for 2nd order Epistasis Detection.}
    \label{fig:algooo}
\end{figure}

We illustrate a high-level view of the circuit in Figure \ref{fig:algooo}. For brevity, we present the circuit for 2nd order, following the logic present in Section \ref{sec:enunciating Epistsais}, for a pair of \glspl{SNP} X and Y. Their content will be matched by a neuron simulating \verb|AND|, with its threshold set to $1$, only spiking if both inputs are $1$. This is done in parallel for every column of the dataset, such that each column is associated with a dedicated \verb|AND| neuron. Its outputs feed a specially designed \gls{POPC} circuit, described in Section \ref{sec:SNN POPC}, which counts the number of spikes from all the \verb|AND| neurons. The result of the \gls{POPC} circuit is a binary spike representation of the relative frequency of the selected \gls{SNP} pair values in the dataset. The output is read on the fly by an external digital controller and stored in host memory.

We embed the dataset within the \gls{SNN} architecture by storing each \gls{SNP}'s mask in dedicated synapses as described in Section \ref{sec:syn stackcs}. We do it by encoding \glspl{SNP} values as firing patterns in an addressable non-destructive stack so that we can reproduce any \glspl{SNP} values by routing a spike to its dedicated stack circuit. All \gls{SNP} pairs are selected by synchronously activating pairs of \glspl{SNP}, sending control input spikes, following a pre-determined activation order. Combinations between a pair's possible values, are generated by activating Y's synapse three times, and sending X's content to a specially designed repeater circuit, described in Section \ref{sec:Repeater circuit}, whose output sequence will match the possible values of X with Y's stack circuit output sequence.

\subsubsection{Hardware-Based Restrictions} 
\label{sec:hardware restrictions}

To ensure hardware implementability, we introduce practical constraints that limit real complexity gains of a direct threshold gate implementation: limiting neuron fan-in/fan-out to constants $F_{in}/F_{out}$; representing currents and thresholds as signed integers with precision $N_{pr}$; representing weights by signed integers with precision $S_{pr}$; and imposing a maximum delay $M_{delay}$ on synapses. These restrictions prevent theoretical mechanisms that allow, for instance, to encode infinite information in neuronal currents \cite{siegelmann1992computational}. For design simplicity, we assume $F_{in} \approx F_{out}$ and $S_{pr} \leq N_{pr}$. These constraints are consistent with state-of-the-art systems such as Loihi \cite{Loihi1} and Akida \cite{vanarse2019hardwareAKIDA}, ensuring the model adheres to limitations of the \gls{NoC} for spike distribution. Table \ref{tab:hardw limits} summarizes the hardware restrictions introduced.

\begin{table}[t]
\caption{Hardware Limit Constants}
\label{tab:hardw limits}
\centering
\begin{tabular}{|l|l|}
\hline
Name          & Description                                              \\ \hline
$S_{pr} ~$      & Signed integer synaptic weight precision                                \\
$N_{pr} ~$      & Signed integer neuron current and threshold precision                 \\
$M_{delay} ~$ & Max. synaptic transmission delay                            \\
$F_{in} ~$    & Max. number of synapses connected \textbf{to} a neuron   \\
$F_{out} ~ $   & Max. number of synapses connected \textbf{from} a neuron \\ \hline
\end{tabular}
\end{table}

For a circuit with limited fan-in that depends on all $n$ inputs (i.e. all inputs are connected to the output neurons), one should not target constant depth as the best attainable is $O(log_{F_{in}}n)$. This notion can be formalized by considering a bounded fan-in threshold gate circuit computing a function $f:\{0,1\}^n \to \{0,1\}$. The output gate connects to at most $F_{in}$ gates at the previous level, each connecting to at most $F_{in}$ gates of their own. To obtain optimal depth, a tree is built with branches of size $F_{in}$, obtaining $O(log_{F_{in}}n)$ depth \cite{TC_bounded_fan}.

\section{Data Encoding and Processing}
\label{sec:Seq encoding and Gen}

We propose embedding the dataset in \gls{SNN}'s hardware by encoding values as firing patterns in an addressable non-destructive way in stack circuits, allowing reproduction by routing a spike to activate the pattern firing. We also propose a repeater circuit to generate combinations between \glspl{SNP} to feed the \gls{POPC} circuit. 

\subsection{Synaptic Stack Encoding}
\label{sec:syn stackcs}

\begin{figure}[t]
    \centering
    \includegraphics[width=0.6\linewidth]{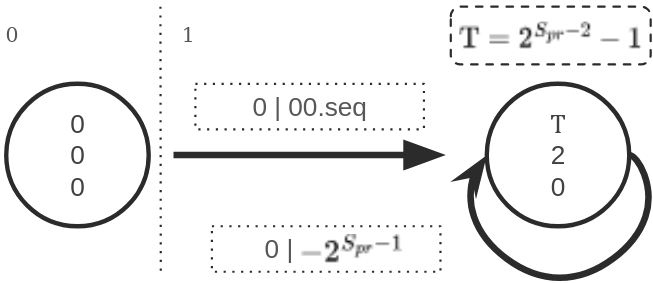}
    \caption{Pattern memory circuit}
    \label{fig:memory}
\end{figure}

We propose the circuit shown in Figure \ref{fig:memory} to create a mechanism for neurons to produce specific firing patterns on demand. We can view a firing pattern of a neuron as a bit-string stack that is \textit{pop} every timestep, the top bit determining whether the neuron fires (1) or remains idle (0). The bit-string is stored starting in the second most significant bit of the synaptic weight, with the sign and the most significant bit set to $0$. The level 0 neuron acts as a trigger, delivering the content of the synapse to the level 1 neuron's current, which will fire according to the encoded sequence.

By setting the threshold to $T= 2^{S_{pr}-2}-1$, the neuron will automatically fire according to the top bit of the stack. The neuron's current undergoes a left shift operation at each timestep by setting the leakage $m$ to $2$. A self-connected inhibitory synapse with weight set to $-2^{S_{pr}-1}$, cancels out the top bit. This way, the sequence is reproduced one bit at a time. The maximum encoded sequence length is $S_{pr}-2$.

\subsection{Repeater Circuit}
\label{sec:Repeater circuit}

\begin{figure}[t]
    \centering
    \includegraphics[width=1\linewidth]{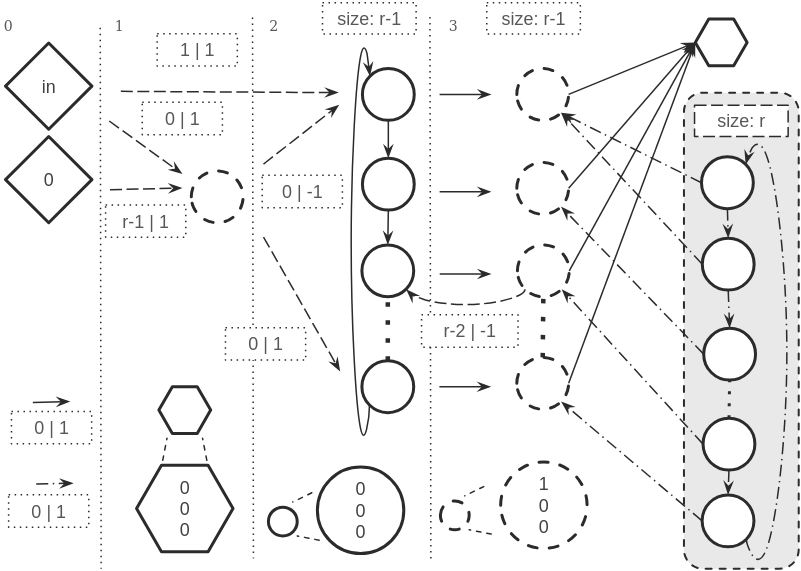}
    \caption{Repeater Circuit}
    \label{fig:Repeat}
\end{figure}

To generate combinations, we propose a repeater circuit, presented in Figure \ref{fig:Repeat}, that transforms the firing pattern generated by its \textit{in} neuron (as described in Section \ref{sec:syn stackcs}) by reproducing each bit of the $r$-bit input sequence exactly $r$ times in the output neuron's firing pattern. For example, if the input neuron fires according to the pattern $010$, the output neuron will produce the pattern $000111000$, where each bit from the original sequence is repeated three times. The key mechanism involves temporarily storing the sequence in cyclically connected neurons acting as spike repeaters. At each timestep, consecutive neurons in the cycle fire according to consecutive bits of the input sequence. The cyclic connection scheme causes each bit to propagate around the cycle, which causes consecutive neurons to fire according to that bit at each timestep. By tracking which neuron corresponds to each bit position at every timestep, we can selectively route specific bits to the output neuron. 

The circuit consists of four distinct layers. The level $2$ neurons are precisely the cyclically connected spike repeater neurons mentioned above, which temporarily store the input sequence and enable the bit propagation. The layer is composed of only $r-1$ neurons, thus not having space for the last bit. The level $1$ neuron serves solely to correctly position bit $r$ once the 1st bit isn't needed. With its threshold set to $1$, it spikes upon receiving a programmed spike at timestep $r$, subsequently inhibiting the sequence input and positioning the last bit at the first bit's location after $r$ timesteps. Each level $2$ neuron connects to a corresponding level $3$ neuron that functions as a gate, firing only when it simultaneously receives input from one of the neurons in the shaded box. The shaded box contains $r$ neurons that activate sequentially, each exciting its adjacent neuron. This activation sequence aligns with bit positions in level $2$. Two shaded box neurons are connected to the same gate, thus, every $r$ timesteps, the gate opening sequence is shifted to align with the preceding bit cycling around layer 2. One specific gate is also responsible for cleaning, by having an inhibitory synapse connected to level $2$, eliminating a bit after it completes one cycle, as the synaptic delay is $r-2$. The spikes generated by the gate neurons are transmitted to the output neuron.

\subsubsection{Complexity Analysis}
As a recurrent network, this circuit's performance is measured by the overhead between receiving and repeating a sequence. For this circuit, it corresponds to the length of the minimal path from input to output. All the bits follow in consecutive timesteps. Therefore, the circuit introduces a three-timestep initial delay overhead that does not depend on the size of the sequence. It introduces $O(1)$ overhead for a minimal $O(r^2)$ runtime. In terms of space, we note that it is sparsely connected as it uses only $O(r)$ neurons and synapses. Energy-wise, the last bit will be spiking around layer 2 $r^2$ times, leading the circuit to use a maximum cap of $O(r^2)$ energy. Lastly, unlike the other circuits, the delays introduced scale with $O(r)$. If $r > M_{\text{delay}}$, additional neurons can be inserted to extend delays without increasing the sequence output overhead.

\section{Binary Population Count}
\label{sec:POPC_general}

We propose a logarithmic-depth circuit for generating the result in unsigned binary format the \gls{POPC} of a bit-string, considering the hardware restrictions presented in Section \ref{sec:hardware restrictions}. We start by developing separately two unrestricted threshold gate circuits, one for \gls{POPC} and other for sum of a binary array, to then compose them with hardware restrictions.

\subsection{Constant Depth Threshold Gates Population Count}
\label{sec:POPC}

\begin{figure}[t]
    \centering
    \includegraphics[width=1\linewidth]{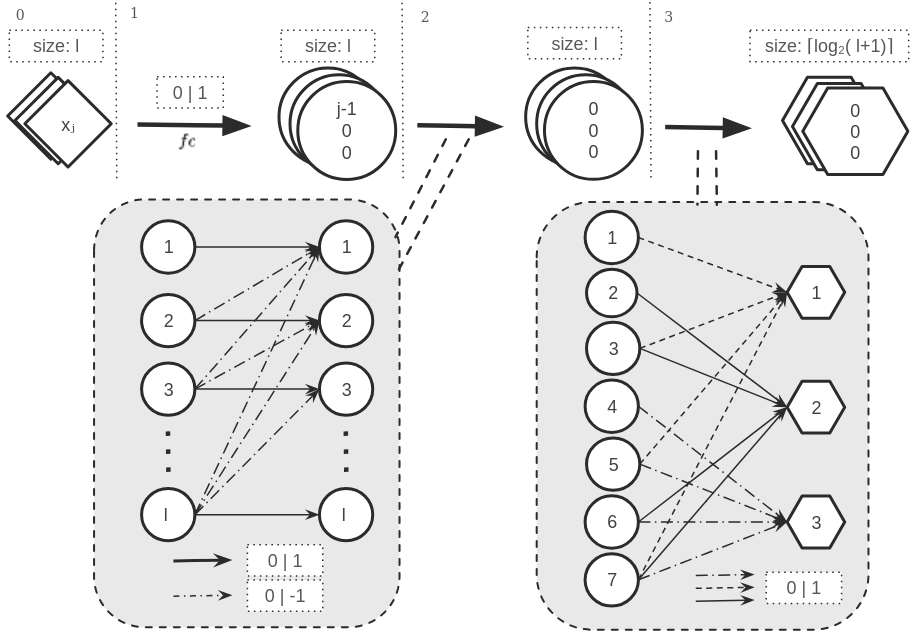}
    \caption{Threshold gate circuit to perform binary \gls{POPC}.}
    \label{fig:POPC}
\end{figure}

The proposed threshold gate circuit \gls{POPC} circuit is schematized in Figure \ref{fig:POPC}. We use $l$ for the input size to ease the analysis of circuit composition. The neurons on each layer are indexed with $j \geq 1$. The circuit computes the number of input spikes into a binary representation using three levels of depth. Let $(y_1, y_2, \dots, y_l)$ be a bit-string encoded in the spikes of the input layer, where a neuron fires if its corresponding bit is $1$ and remains silent if the bit is $0$. First, the number of spikes in the input layer are counted and represented by a one-hot encoding of the possible sum count values, i.e. each neuron is represents a number with only one neuron allowed to spike. This enables a direct translation into an unsigned binary representation of the result.

The first level of the circuit counts the number of input spikes. It is composed of $l$ neurons with increasing thresholds $t \in \{0, 1, 2, \dots, l-1\}$, such that neuron $j$ has threshold $j-1$. Each depth 1 neuron is connected to every input with weight-one synapses. When the input layer spikes, the current will be set to the sum of the bits. The number of neurons to spike in the next timestep is thus exactly $\sum_{i=1}^l y_i$.

The second level generates a one-hot encoding of possible $l$ count values. It consists of $l$ neurons where each neuron with index $j$ receives excitatory connections from the depth $1$ neuron with equal index and inhibitory connections from all bigger-indexed neurons, as specified in the left shaded box. This ensures that exactly one neuron in the second level will spike, representing the precise count of active inputs.

The third level consists of $\lceil \log_2(l+1) \rceil$ neurons, each representing a bit of the output binary representation. Each second-level neuron is connected only to the output neurons for which the corresponding bit in the binary representation is $1$. The neuron representing bit $i$ is thus connected to every other set of consecutive $2^i$ numbers represented at level $2$. Given a number $x \in \mathbb{N}$ and $x_i$ being the $i$th bit of $x$'s binary representation, the following relation holds:
\begin{equation}
\label{eq:conditiononnn}
    \begin{cases}
        0 ~ \leq ~ x < 2^i &\text{ mod } 2^{i+1}  ~ ~ ~  \text{ \it iff} ~ ~ ~  x_i = 0 \\
        2^i \leq x < 2^{i+1} &\text{ mod } 2^{i+1} ~ ~ ~ \text{ \it iff} ~ ~ ~  x_i = 1
    \end{cases}
\end{equation}
showing that there is a well-defined sequence of numbers for the value of the bits, that is followed by the connection scheme. Figure \ref{fig:POPC} illustrates it up to the third output bit.

\subsubsection{Complexity Analysis}
As can be inferred by Figure \ref{fig:POPC}, the circuit has $O(1)$ time complexity because of its constant depth. In terms of space, it has 
$O(l)$ neurons. Further, the number of synapses 
amounts to $O(l^2)$ has the input layer is fully connected with the depth 1 layer. The maximum in-degree as well as out-degree is $l$. There are no synaptic delays. The energy complexity depends on the actual value of the input and grows with the encoded number. It is bounded $O(l)$ spikes.

\subsection{Constant Depth Binary Sum}
\label{sec: binary sum}

\begin{figure}[t]
    \centering
    \includegraphics[width=0.95\linewidth]{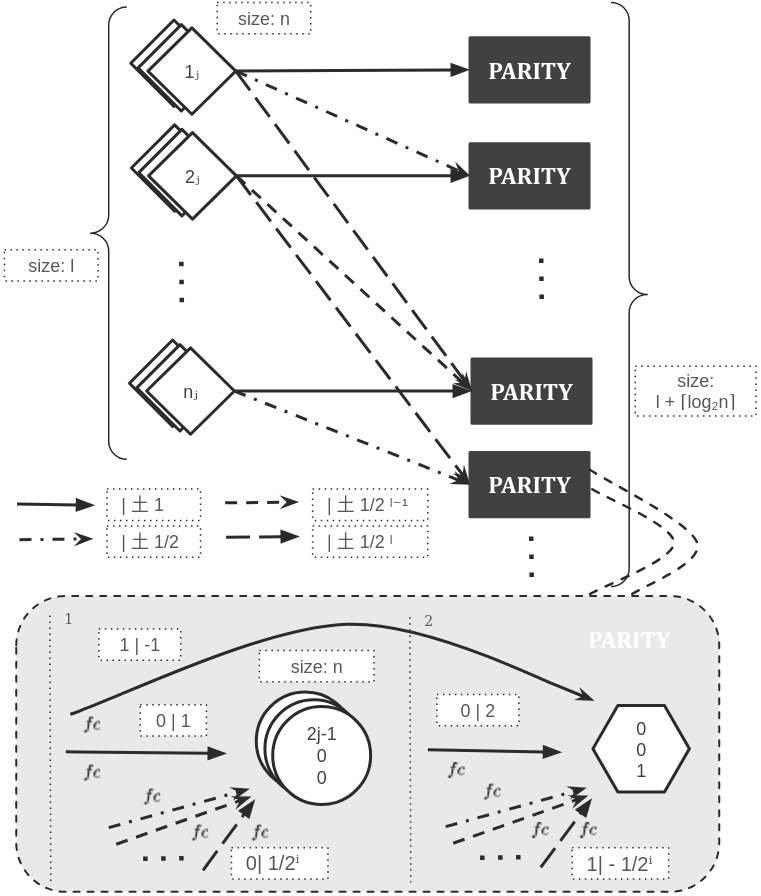}
    \caption{Threshold gate binary sum circuit.}
    \label{fig:binary_sum}
\end{figure}

Limited fan-in hardware restrictions, impose that \gls{POPC} circuit input size must be limited. To extend it, we compute \gls{POPC} separately for parts of the input and then sum the results. We propose thus a general threshold gate circuit to sum an array of $n$ binary numbers up to $l$ bits each. The idea exploits the fact that the bits of the result are given by checking the parity of the bits at each position along with the carry bits induced by less significant ones. 

The circuit is schematized in Figure \ref{fig:binary_sum}. Let $(y^{(j)}_1, y^{(j)}_2 \dots, y^{(j)}_{l})$ be the $j$-th binary input, with $y^{(j)}_i$ being its $i$-th bit, and $(z_1, z_2, \dots, z_{l+\lceil log_2(n) \rceil})$ be the output. The input is organized in parallel layers by input bit significancy; i.e. the input layer $2_j$ corresponds to the bit-string of all the second bits of the inputs. The output bits are calculated each by a parity circuit (Section \ref{sec:Constant Depth XOR}), modified to incorporate input corresponding to the carry bits. Each of the bit's input layer, is connected to the 
parity circuits corresponding to the bigger indexed output bits with synaptic weight set to $1/2^{k-i}$ where $k$ is the index of the output bit and $i$ is the index of the input bit layer. 

The modified parity circuit is represented in the shaded box. It follows the same principle as in Section \ref{sec:Constant Depth XOR}, with adapted sizes. The induced current in layer $1$ will be:
\begin{equation}
\label{eq:bits}
    \sum_{i=1}^{k}\sum_{j=1}^{n} \frac{y_i^{(j)}}{2^{k-i}},
\end{equation}
that is the same as if following bit string:
\begin{equation}
    un \left\lfloor \sum_{i=1}^{k-1} \sum_{j=1}^{n} \frac{y_i^{(j)}}{2^{k-i}} \right\rfloor.(y_k^{(1)}y_k^{(2)}\dots y_k^{(n)}),
\end{equation}
composed of the carry bits concatenated with the input bits at position $k$, where $un\lfloor\cdot\rfloor$ in the unary bit-string corresponding to the integer floor, was the input already, as the layer's threshold are integers. The same rationale is true for the depth $2$ output neuron. We conclude than that the output bit of the schematized circuit in position $k$ is the following:
\begin{equation}
\label{eq:holy grail}
    z_k =  \left( \bigoplus un \left\lfloor \sum_{i=0}^{k} \sum_{j=0}^{n-1} \frac{y_i^{(j)}}{2^{k-i}} \right\rfloor ~ \right) \oplus y_k^{(0)} \oplus ~ ... ~ \oplus  y_k^{(n)}
\end{equation}
It remains to show that \eqref{eq:bits} is not miscalculating the number of carry bits. The true number, for $k >i$ is equal to:
\begin{equation}
\label{eq:actual bits}
    \sum_{i=1}^{k}\left\lfloor \frac{\sum_{j=1}^n y_i^{(j)}}{2^{k-i}}\right\rfloor.
\end{equation} The difference between \eqref{eq:actual bits} and \eqref{eq:bits} is less than 1:
\begin{equation}
    \sum_{i=1}^{k}\sum_{j=1}^{n} \frac{y_i^{(j)}}{2^{k-i}} -
    \sum_{i=1}^{k}\left\lfloor \frac{\sum_{j=1}^n y_i^{(j)}}{2^{k-i}}\right\rfloor < \sum_{i=1}^{k} \frac{1}{2^i} < 1
\end{equation}
showing the correction of the circuit.

\subsubsection{Complexity Analysis}

The parity circuit has constant depth, as $l+\lceil \log_2n\rceil$ parity circuits are needed in parallel, presenting $O(1)$ time complexity. In terms of space, it requires 
$O(ln)$ neurons. To calculate the number of synapses, we notice that each input bit layer is connected to $l + \lceil \log_2 n\rceil$ parity circuits. Therefore, replacing the value in the synapse count in Section \ref{sec:Constant Depth XOR} by  $(l+\lceil \log_2n\rceil )\times n$, multiplied by $l$ input binary numbers, we arrive at the total number of synapses of the circuit
$O(l^3n^2 + l n^2\log^2 n )$. The maximum in-degree is $ \left\lfloor \frac{ 3 (l +\lceil \log_2n\rceil) n}{2} \right \rfloor$, and the maximum out-degree is $(l+\lceil \log_2n\rceil) n +1$. The maximum synaptic delay is 1. The energy depends on the actual value of the input and grows with the number of input spikes. It is bounded by all the inputs spiking, leading to $O(l n)$ spikes.

\subsection{Hardware Restricted Population Count Circuit}
\label{sec:SNN POPC}

\begin{figure}[t]
    \centering
    \includegraphics[width=1\linewidth]{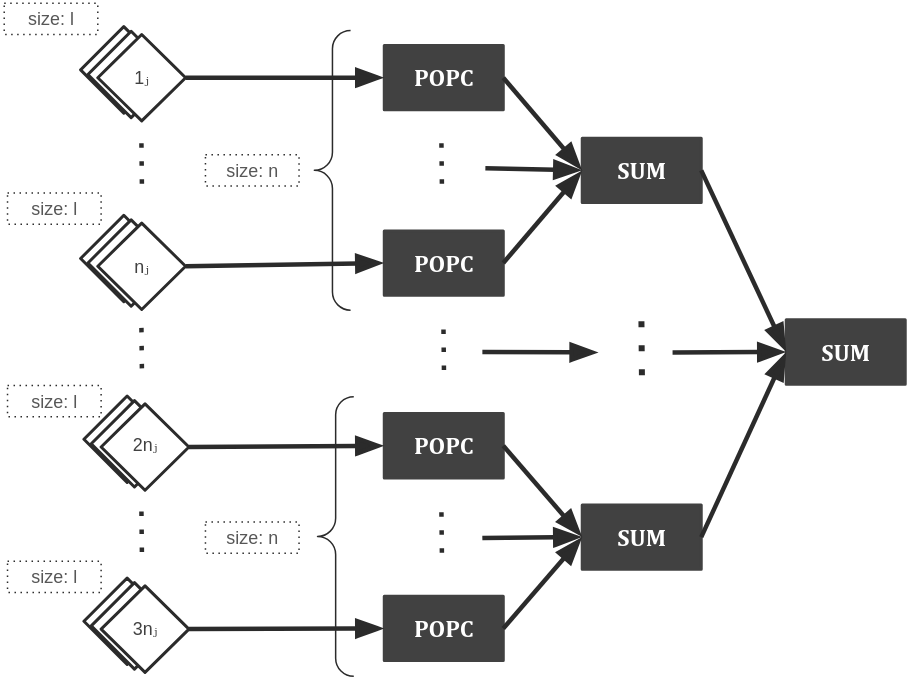}
    \caption{\gls{SNN} \gls{POPC} circuit with hardware restrictions.}
    \label{fig:hard-restric_POPC}
\end{figure}

We propose the circuit present in Figure \ref{fig:hard-restric_POPC} to perform \gls{POPC} with limited fan-in/out and variable precisions. We divide the input into parts of size $l$, and perform \gls{POPC} on each part using the algorithm from Section \ref{sec:POPC}. Using the output from the \gls{POPC} of each part, we proceed to sum the binary results in a descending tree fashion, summing $n$ binary numbers at a time using the circuit from Section \ref{sec: binary sum}. This approach allows us to perform \gls{POPC} of an array of any size $m$. We need to determine the optimal values for $l$ and $n$ in line with restrictions. We have initially $l \leq F_{in}$, because the maximum fan-in of the \gls{POPC} circuit equals the input size. For the sum circuits $l$ will grow automatically to to accommodate bigger binary numbers. Therefore we benefit from minimizing $n$, leading to approximate synaptic count of the sum circuit of $O(l^3)$ and maximum out-degree of $3l \leq F_{in}$.

\subsubsection{Complexity Analysis}

Overall, an input of size $m$ has $m/l$ \gls{POPC} circuits and $O(m)$ circuits sum circuits with at most $\lceil\log_2 m\rceil$ bits. This yields a time complexity of $O(\log m)$. The total number of synapses is $O(m\log^3 m)$ from the binary sums plus $O(m)$ from the \gls{POPC}. The number of neurons is $O(m)$ from the \gls{POPC} plus $O(\log m)$ from the binary sums. The maximum synaptic delay is $1$. In terms of energy, it depends on the size of the input. It is bounded by all inputs of the \gls{POPC} circuit spiking, amounting to $O(m\log m)$.

\subsubsection{Encoding fractional weights}

As stated in Section \ref{sec:hardware restrictions}, synapse weights are encoded as integer values of limited precision. To encode fractions, we scale the threshold of the receiving neurons by $2^{S_{pr}-1}$. Then $\frac{1}{2^k}$ becomes encoded as $2^{S_{pr}-k-1}$. Therefore, the lowest fraction we can encode is $\frac{1}{2^{S_{pr}-2}}$. This gives us the further restriction that $\lceil \log_2 m \rceil \leq S_{pr} - 2$, which implies that $m \leq 2^{S_{pr}-2}$. 

\section{Discussion}
\label{sec:discussion}


The algorithm works in a fully pipelined manner, such that an entry is computed at each timestep. Let the number of \glspl{SNP} be $n$ and the number of samples be $m$. The shortest path from input to output takes $O(\log n)$ timesteps, as the \gls{POPC} circuit (Section \ref{sec:SNN POPC}) has logarithmic depth. Therefore, the total time is $O(\binom{n}{k}+ \log n ) = O(n^k)$, for $k \ll n$.

The circuit is favorable to a low energy implementation, as each memory and combination's algorithm neuron will spike $1/3$ of the time because of the dataset binarization. The \gls{POPC} circuit energy will depend on the frequency of each combination across samples. Still, as there are usually a low number of high-frequency candidates, the overall spiking of the \gls{POPC} circuit will be marginal compared to the combinations generating algorithm. Anyhow, the number of spikes will be limited by those high-frequency candidates introducing an additional $O(log m)$ per combination. This lead to a complexity of $O(n^2m\log m)$, though wasting less energy on average.

It occupies overall $O (mn + m\log^3 m)$ synapses, $O(mn)$ neurons because the space is limited by the size of the dataset encoding (Section \ref{sec:syn stackcs}) and depth of the \gls{POPC}. As $k \ll n$, the space occupied by the repeaters can be neglected. The number of synapses can also be reduced by compacting multiple \gls{SNP}s into each synapse, up to the precision of its weights, though requiring the calculation of unwanted combinations. The approach can be extended to $k$-order interactions by adding repeaters to the synaptic memory output path. For instance, with $k=3$, we would add another repeater in which each bit would be repeated $9$ times instead of $3$.

The algorithms proposed are each interesting on their own. The suggested approach for encoding memories in the network enables the implementation of tailored dynamical systems and control schemes that prevent off-chip spikes. Binary summation and \gls{POPC} circuits can be utilized in a variety of direct computing applications and large dataset processing, as they present complexity polynomial gains when compared with classical counterparts, further allowing for a pipelined execution. A summary of the complexity results obtained is summarized in Table \ref{tab:results}. Variable names are consistent with the naming used in the sections describing the circuits.

\begin{table}[t]
\caption{Summary of Complexity Results}
\label{tab:results}
\centering
\begin{tabular}{|l|c|c|}
\hline
Circuit          & Time            & Energy            \\ \hline
$TC$ Parity        & $O(1)$          & $O(n)$            \\
$TC$ POPC          & $O(1)$          & $O(l)$            \\
$TC$ Sum           & $O(1)$          & $O(ln)$           \\ \hline
POPC             & $O(\log m)$    & $O(m)$            \\
Repeater         & $O(r^2)$        & $O(r)$            \\
Epistasis        & $O(n^k)$        & ~ $O(n^k m \log m)$ ~ \\ \hline
\end{tabular}
\end{table}

\begin{table}[t]
\centering
\begin{tabular}{|l|c|c|}
\hline
Circuit          & Neurons         & Synapses             \\ \hline
$TC$ Parity        & $O(n)$          & $O(n^2)$             \\
$TC$ POPC          & $O(l)$          & $O(l^2)$             \\
$TC$ Sum           & $O(ln)$         & $O(l^3n^2 + l n^2\log^2 n )$    \\ \hline
POPC             & $O(m)$          & $O(m \log^3 m)$        \\
Repeater         & $O(r)$          & $O(r)$               \\
Epistasis        & $O(mn)$         & $O(mn + m \log^3 m)$   \\ \hline
\end{tabular}
\end{table}

\section{Conclusions and Future Work}
\label{sec:conclusions}

In this work, we have proposed a neuromorphic realization of an algorithm for epistasis detection that keeps optimal complexity features while accounting for hardware restrictions. By embedding the dataset in addressable stacks, we feed an efficient $TC^0$ algorithm performing all necessary operations without complexity overhead. While showing theoretical gains, more research is necessary to compare with classical algorithms. The comparison can be theoretically improved by incorporating data movement costs, which vary based on the mapping distance between connected neurons in the chip's cores, which we leave as future work.

The algorithm can be amenable to analog implementation. Stack memory operations can be reproduced in the analog domain by representing bits as a subset of base-4, in the Cantor 4 set, leaving gaps between encoded values. This approach allows implementations to tolerate natural errors and hardware noise without requiring high precision \cite{siegelmann1992computational}. However, the formal accounting of the overhead introduced remains unclear and varies between technologies.

As neuromorphic hardware scales, efforts like this prepare the field to tackle complex real-world problems where neuromorphic advantages can be explored across diverse applications. The complexity gains demonstrated by tackling epistasis detection highlight the potential suitability of large neuromorphic engines for high-performance tasks with lower energy footprints and asymptotic efficiency.


\section*{Acknowledgment}
This work was supported by Fundação para a Ciência e a Tecnologia, Portugal, EuroHPC Joint Undertaking and European Union HE Research and Innovation programme through UIDB/50021/2020 project and GAs No 101143931 (POP3) and 101092877 (SYCLOPS).


\bibliographystyle{IEEEtran}
\bibliography{IEEEabrv,Organised_Neruomorphic}

\end{document}